\documentclass[10pt, conference, compsocconf]{IEEEtran}
\usepackage{amsmath}
\usepackage{graphicx}
\usepackage{hyperref}
\ifCLASSINFOpdf
\else
\fi
\begin{document}
%
\title{Cooperative Game Theoretic Bid Optimizer for Sponsored Search Auctions}
\author{\IEEEauthorblockN{Sriram Somanchi}
\IEEEauthorblockA{Electronic Commerce Laboratory\\
Computer Science and Automation\\
Indian Institute Science, Bangalore \\
Email: somanchi@csa.iisc.ernet.in}
\and
\IEEEauthorblockN{Chaitanya Nittala}
\IEEEauthorblockA{Electronic Commerce Laboratory\\
Computer Science and Automation\\
Indian Institute Science, Bangalore \\
Email: chytu@csa.iisc.ernet.in}
\and
\IEEEauthorblockN{Narahari Yadati}
\IEEEauthorblockA{Electronic Commerce Laboratory\\
Computer Science and Automation\\
Indian Institute Science, Bangalore \\
Email: hari@csa.iisc.ernet.in}
}
%

\maketitle

\begin{abstract}

In this paper, we propose a bid optimizer for sponsored keyword search auctions which leads to better retention of advertisers by yielding attractive utilities to the advertisers without decreasing the revenue to the search engine. The bid optimizer is positioned as a key value added tool the search engine provides to the advertisers.
The proposed bid optimizer algorithm transforms the reported values of the advertisers for a keyword into a correlated bid profile using many ideas from cooperative game theory. The algorithm is based on a characteristic form game involving the search engine and the advertisers. Ideas from
Nash bargaining theory are used in formulating the characteristic form game to provide for a fair share of surplus among the players involved.
The algorithm then computes the nucleolus of the characteristic form game since we find that the nucleolus is an apt way of allocating the gains of cooperation among the search engine and the advertisers. The algorithm next transforms  the nucleolus into a correlated bid profile using a linear programming formulation. This bid profile is  input to a standard generalized second price mechanism (GSP) for  determining the  allocation of sponsored slots and the prices to be be paid by the winners.
The correlated bid profile that we determine is a locally  envy-free equilibrium and also a correlated equilibrium of the underlying game. Through detailed simulation experiments, we show that the proposed bid optimizer retains more customers than a plain GSP mechanism and also yields better long-run utilities to the search engine and the advertisers.
\end{abstract}

\begin{IEEEkeywords}
Bid Optimizer,
Sponsored Search,
Cooperative Game Theory,
Nash bargaining,
Nucleolus.
\end{IEEEkeywords}

\IEEEpeerreviewmaketitle

\section{Introduction}
Sponsored search auctions have been studied extensively in the recent years due to the advent of targeted advertising and its role in generating large revenues.
 With a huge
competition in providing the sponsored search links, the search
engines face an imminent problem which can be called as the
\emph{retention problem}. If an advertiser (or alternatively bidder) does not get satisfied
because of not getting the right number of clicks or the
anticipated payoff, he could drop out of the auction and try
sponsored links at a different search engine.

\subsection{Motivation: Retention of Advertisers in Sponsored Search Auctions}
Our motivation to study the retention problem is driven by the compulsions faced by both the search engine and the advertisers.

From the advertisers' perspective, choosing their
maximum-willingness-to-pay such that they get an attractive slot
subject to their budget constraints is a challenging problem. The
search engines can use various mechanisms for the sponsored search
auction as described in ~\cite{agt:noam,monograph} but the most
popular mechanism is the generalized second price (GSP) auction
since it is simple and yields better revenue to the search engine.
In the most simple version of GSP, where there are $k$ slots and
$n$ advertisers (for simplicity assume $k \leq n$), the allocation
and payment rule are as follows. The allocation rule is that $n$
advertisers are ranked in descending order based on their bids,
with ties broken appropriately, and top $k$ advertisers'
advertisements are displayed. The payment rule is that every
advertiser needs to pay bid amount of the advertiser who is just
below his slot and last advertiser is charged the highest bid that
has not won any slot. If the non-truthful GSP
auction~\cite{EdelOstros} is used by the search engine, the
bidders will have an incentive to shade their bids. The bidders would not want to use complicated and computationally intensive bidding strategies as the bidding process is done many times (typically thousands of times) in a day. These advertisers generally build their own
software agents or employ third party software agents, which
adjust and readjust the bid values on behalf of these advertisers.
The bidders typically specify their maximum willingness-to-pay for
their keywords for any given day. Hence each keyword has a
specific set of bidders bidding on it for the whole day. This
scenario constitutes a repeated game between all the bidders
bidding for that keyword. In this game, bidders who cannot plan
their budget effectively may experience less utilities and thus
may drop out of the system.

We now turn to the search engine's perspective of the retention problem. When the bidders try to know each others' valuations  by submitting and resubmitting bids, they may find a set of strategy profiles which may yield all of them better payoffs. This may lead to collusion among the bidders. Folk theorems~\cite{mwg} suggest that players may be able to increase individual profits by colluding thereby decreasing the  search engine's revenue. Even though the bidders in the keyword auctions are competitors, this collusion against the search engine could be stable. Vorobeychik and Reeves~\cite{motivation} studied this phenomenon   and illustrated a particular collusive strategy which is better for all the bidders (hence worst for the search engine) and can be sustainable over a range of settings. Feng and Zhang  ~\cite{feng} showed that dynamic price competition between competing advertisers can lead to collusion among them. However, in this dynamic scenario, when the discounted payoffs of the bidders under the collusive strategy are considered, the stability of collusion depends inversely on the number of bidders~\cite{mwg}. That is, the lower the number of bidders in the system, the higher is the stability of the collusion. This motivates us to study the bidder retention problem for the search engine.

Also, due to exponential growth in the space of online advertising
and intense competition among the search engine companies, the
switching cost for the advertisers to change from one search
engine to another is almost zero~\cite{agt:noam}. Hence, it is
imperative for the search engine companies to retain their
advertisers to safeguard their market share. Driven by this, the
search engine companies have introduced many value added tools,
such as bid optimizer, to maximize the bang-per-buck for the
bidders. In what follows, we describe the bid optimizer's role in
solving the retention problem.

\subsection{Bid Optimizers}
A \textit{bid optimizer} is a software agent provided by the search
engine in order to assist the advertisers. The bidders are required to
provide to the bid optimizer a target budget for the day
and a maximum willingness-to-pay. Bid optimizers,
currently provided by the search engines, promise to maximize the
revenue of advertisers by adjusting the bid amount in each round
of the auction based on the projected keyword traffic and
remaining budget.

It can be seen that the decisions made by the bid optimizer are
crucial to both the search engine and the set of advertisers, who
choose to use the bid optimizer.
Hence, the objective of a typical bid optimizer is to
strike a balance between reduction in revenue of the search engine
company versus increase the retention of advertisers. This
objective is achieved by providing enhanced utilities to the
advertisers, thus ensuring retention of customers, thereby sustaining high levels of revenue to the
search engine company in the long run. Designing such intelligent bid optimizers is
the subject of this paper.

There are some problems involved in designing bid optimizers.

\begin{enumerate}
 \item For the search engine, maximizing its short-term revenue (that is, its payoff in a one-shot game) seems to be a viable option. But here, the lower valuation bidders are denied slots due to allocative efficiency concerns. For the bidders, as shown by Cary \textit{et al}~\cite{cary}, where all the high valuation bidders use a particular greedy strategy, it has been proved that none of the bidders except the top $k$ bidders get the slots after a certain number of rounds of the auction. The above phenomenon can permanently drive away low valuation bidders from the search engine.

 \item Dropping out of the search engine to get better utilities in another search engine is a possible option for the bidders. The low valuation bidders drop out after not getting slots for a certain period of time. The higher valuation bidders can observe this trend and shade their maximum-willingness-to-pay or collude to get better utilities. This may result in the search engine losing  revenue. This is a threat to the search engine from  the bidders. However, if a large number of bidders remain in the system, the collusion is not stable. The intuition for this is that, high valuation bidders cannot reduce their bids sharply, since they will have the fear of undercutting the lower valuation bidders present in the system and thus losing out on their slots.
\end{enumerate}

Hence, we propose that retaining more number of bidders solves all the problems discussed above. The dependence of the search engine and the bidders on each other for mutual benefit motivates us to use a cooperative approach in general. The above threat model naturally directs us towards using a Nash bargaining model in particular. Our solution can be seen as associating the bid optimizer to a keyword rather than bidders as done by the existing bid optimizers. The overall model of the bid optimizer is depicted in Figure \ref{fig:blk1}.

\subsection{Contributions and Outline of the Paper}

\input{epsf}
\begin{figure*}[!hbtp]
 \centering
\epsfysize=9.5cm
\epsfxsize=14.5cm
\epsfbox{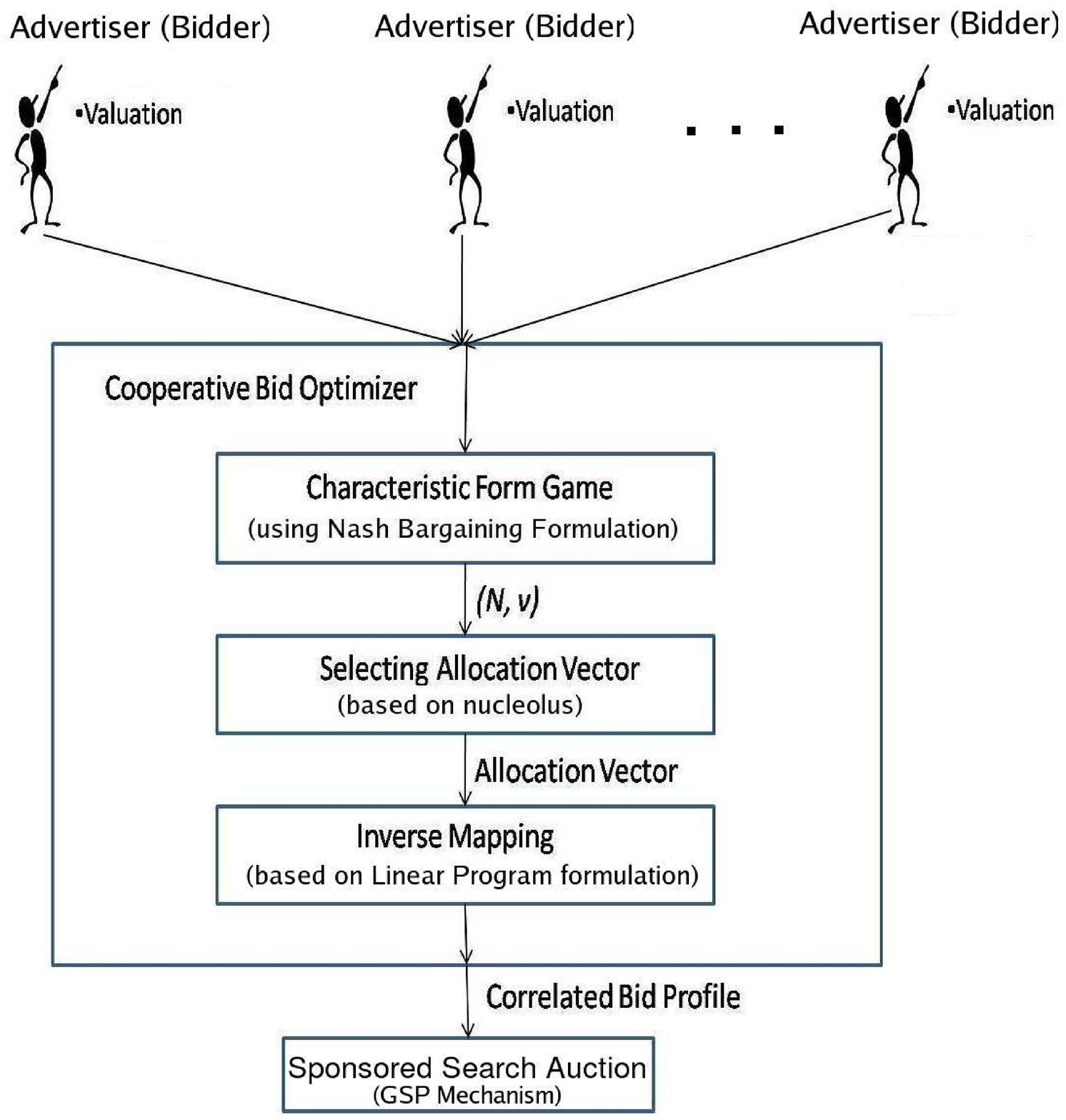}
 \caption{Proposed Bid Optimizer}
 \label{fig:blk1}
\end{figure*}

In this paper, we propose a bid optimizer that uses many ideas from cooperative game theory. The bid optimizer is shown in Figure \ref{fig:blk1}.

\begin{itemize}
 \item The inputs to the bid optimizer are the willingness-to-pay values (or valuations) of the bidders.
 \item The output of the bid optimizer is a correlated bid profile, which, when input to a standard GSP auction mechanism, yields utilities to the search engine and the advertisers satisfying the goals set forth in the paper.
 \item The bid optimizer first formulates a characteristic form game involving the search engine and the advertisers. The value for each coalition is defined based on a novel Nash Bargaining formulation with the search engine as one player and a virtual player aggregating all advertisers in that coalition as the other player. The idea of using Nash Bargaining is to ensure a fair share for the search engine and the advertisers.
 \item The nucleolus of the above characteristic form game is selected as the utility profile for the search engine and the advertisers. The choice of nucleolus is based on key considerations such as, bidder retention, stability, and efficiency.
 \item The utility profile represented by the nucleolus is mapped to a correlated bid profile that satisfies individual rationality, retention, stability and efficiency. A linear programming based algorithm is suggested for this purpose.
\end{itemize}

We carry out experiments to demonstrate the viability and efficacy of the proposed bid optimizer. We show, using a credible bidder drop out model, that the proposed bid optimizer has excellent bidder retention properties and also yields higher long-run revenues to the search engine, when compared to the plain GSP mechanism.

The outline of the paper is as follows. Section \ref{MODEL} presents the details of the bid optimizer and introduces the model. In Section \ref{NBFORMULATION},  we present a bid optimization algorithm which uses the Nash Bargaining approach for  ensuring the retention of bidders in the system. We then map this fair share for the aggregated bidder  to a correlated bid profile in Section \ref{CorrBidProfileMapping}.  We analyze the properties of our method in Section \ref{ANALYSIS}. We present our experimental results in Section \ref{EXPERIMENTS} and conclude the paper in Section \ref{SUMMARY}.

\section{Our Approach to Bid Optimization}
In this section, we present our algorithm for bid optimization. The algorithm can be divided into three phases as shown in Figure \ref{fig:blk1}: (1) Characteristic form game definition using Nash bargaining, (2) Computing the utility vectors for the players and (3) Inverse mapping of the utility vector into a correlated bid profile. These are discussed in the following sections. The notation in the remainder of the paper is presented in Table \ref{tb:Notation}.

\subsection{Characteristic Form Game}

\subsubsection{The Model} \label{MODEL}
The sponsored search auction scenario we consider has $n$ bidders
competing for $k$ slots of a keyword. We assume that the
probability that a bidder $i$ gets clicked  on the $j$th slot (or
the click-through rate $CTR_{ij}$) is independent of the bidder
$i$,
that is, $CTR_{ij} = \beta_j$ and we also assume that $ \beta_1
\geq \beta_2 \geq \ldots \geq \beta_k$. Each bidder $i$ specifies
his maximum willingness-to-pay $ \overline{s}_{i} $ to the bid
optimizer. The bid optimizer takes as input all the $
\overline{s}_{i} $'s of the bidders and suggests them a correlated
bid profile. This bid optimization algorithm needs to be invoked
only when the number of bidders in the system or their
willingness-to-pay change. We also assume that the bid of the
player $i$ could be any real number in $[0,\overline{s}_{i}]$.

Given the above model, we define a bargaining problem
\footnote{Refer Appendix for the definition of Nash bargaining
problem} between the search engine and the aggregated bidder and
analyze its properties which will help us in formulating a
characteristic form game.

\begin{center}
\begin{table}
\centering
\begin{tabular}{|l|l|}\hline

\textbf{Notation} & \textbf{Explanation}\\
\hline \hline $ A $ & Auctioneer\\ \hline $ B $ & Aggregated
bidder\\ \hline $ n $ & Total number of players \\ \hline $ k $ &
Total number of slots. We assume $k < n$ \\ \hline $ N $ & Set of
bidders $\lbrace 1, 2, \ldots, n \rbrace$\\ \hline $ K $ & Set of
slots $\lbrace 1, 2, \ldots, k \rbrace$\\ \hline $
\overline{s}_{i} $ & Maximum willingness-to-pay of advertiser $i$
\\ \hline $ S_i $ & Strategy set of bidder $i, \>\>
[0,\overline{s}_{i}] $ \\ \hline $ S $ & Set of all bid profiles
$S_1 \times S_2 \times \ldots \times S_n$ \\ \hline $ s $ & Bid
profile $ (s_1, s_2, \ldots, s_n) \in S $ \\ \hline $ u_i(s) $ &
Utility of bidder $i$ on bid profile $s$ \\ \hline $ U_A(s) $ &
Utility of the auctioneer in the  \\ \hline $\>$        & Nash
bargaining formulation for  bid profile $s$\\ \hline $ U_B(s) $ &
Utility of the aggregated bidder in the \\ \hline $\>$        &
Nash bargaining formulation for  bid profile $s$\\ \hline $
\overline{U}_A $ & $\max_{  s \in S} U_A(s)$\\ \hline $
\overline{U}_B $ & $\max_{ s \in S} U_B(s)$\\ \hline
$ \beta_j $ & Click through rate of any bidder in the $j^{th}$ slot\\

\hline
\end{tabular}\\
\caption{Notation}
\label{tb:Notation}
\end{table}             \end{center}

\subsubsection{Characterization of the Nash Bargaining Solution}\label{NBS}
The motivation for a cooperative approach is the dependence of the
search engine and bidders on each other for their mutual benefit.
Given this, the motivation behind choosing a bargaining approach
is that the amount of short-term loss (or in other words, the
investment of the search engine) for the auctioneer should be
chosen based on the bidders present in the system. The Nash
bargaining approach provides a framework for this amount to be
chosen by the search engine by considering all the bidders as one
aggregate agent whose bargaining power depends on all the maximum
willingness-to-pay of all the bidders present in the system.

The utility of the aggregated bidder is the sum of the utilities
of all the bidders over all possible allocations of slots
(outcomes). Now, the bargaining utility space becomes the two
dimensional Cartesian space which consists of the utility of
auctioneer on one axis and the aggregate bidder's utility on the
other axis. Hence a bargaining solution on this space provides a
good compromise for the search engine from its maximum possible
revenue and thus gives the required investment of the search
engine.

The bargaining space is defined in two dimensional Cartesian
space, with utility of auctioneer $U_A(s)$ along the $x-$axis and
the utility of aggregated bidder $U_B(s) = \sum_{i=1}^{n} u_i(s)$
along the $y-$axis. Let $\overline{U}_A$ and $\overline{U}_B$ be
the maximum possible utilities of the auctioneer and the
aggregated bidder respectively. It can be clearly seen that the
value $\overline{U}_A$ is attained for the bid profile $ s =
(\overline{s}_1, \overline{s}_2, \ldots , \overline{s}_n)$ for
which the corresponding $ U_B(\overline{s}) = \sum_{i=1}^{n}
\left( \sum_{j=1}^{k} \beta_j y_{ij}(\overline{s}) \right)
(\overline{s}_i - \overline{s}_{i+1})  = U_{B}^{'}$ (say).
Similarly, the bid profile $s= (0, \ldots , 0)$ yields the utility
pair $(0, \overline{U}_B)$. Since it is theoretically possible
that all the bidders can collude and bid $(0, 0, \ldots, 0)$, we
choose the point $(0,0)$ in this Nash bargaining space as the
disagreement point. Ramakrishnan \textit{et. al} studied this
problem in ~\cite{ramkey:paper} and characterized the solution $
(U_{A}^{*}, U_{B}^{*})$ to this Nash bargaining(NBS) as
 $$ (U_{A}^{*}, U_{B}^{*})  = (\overline{U}_A ,U_B^{'})  \>\>\> if \> \overline{U}_A \leq \frac{\overline{U}_B}{2}$$
 $$ \hspace{1.5 cm}        =  \left(\frac{\overline{U}_B}{2}, \frac{\overline{U}_B}{2} \right)  \>\>\> otherwise $$

\subsubsection{Definition of the Characteristic Form Game} \label{NBFORMULATION}
We use the above model to define Nash bargaining solution $NBS(N) = U_{A}^{*} + U_{B}^{*}$ where $N$ is the set of bidders participating in the auction.

Let $ N = \lbrace 1, 2, \ldots, n \rbrace$ be the set of all bidders and let $0$ represent the search engine. The characteristic form game $\nu : 2^{N \cup \lbrace 0 \rbrace}\rightarrow \Re $ for each coalition $C \subseteq N \cup \lbrace 0 \rbrace $ is now defined as
$$ \nu(C) =  NBS(C) \>\>\> if \> 0 \in C $$
$$ \hspace{0.25 cm} = 0  \hspace{0.25 cm} otherwise \hspace{0.34 cm}  $$
where $NBS(C)$ is defined as above.
If the search engine is not a part of the coalition, its worth is zero since the players cannot gain anything without the search engine displaying their ads. Otherwise, we associate the sum of utilities in the corresponding Nash bargaining bid profile for that coalition with the search engine as the worth of each coalition. This characteristic function $\nu$ defines the bargaining power of each coalition with the search engine.

\subsection{Computing a Utility Vector for the Players: Use of Nucleolus} \label{IdealProperties}
Since there is an aggregation of the bidders' revenue taking place
in the NBS, we map the utility of the aggregated bidder in the
Nash bargaining solution to a correlated bid profile. The NBS
gives an aggregate amount of investment the search engine has to
make on all the bidders. This investment increases the utility of
the aggregated bidder. This utility has to be distributed to the
bidders in a way that our goal of retention is reached. Ideally,
we would like the allocation to have the following properties.
\begin{itemize}
 \item The bidders must not have incentive for not participating in the bid optimizer (individual rationality-IR).
 \item It must retain as many bidders as possible(retention).
 \item The bidders must not have the incentive to shade their maximum willingness to pay (incentive compatibility).
 \item It should be stable both in the one-shot game of GSP and in the cooperative analysis (stability).
 \item It should divide the entire  worth of the grand coalition  among all the bidders (efficiency).
\end{itemize}
 There are several solution concepts in cooperative game
 theory that one could employ here,for example, the core, the Shapley value, the nucleolus, etc.
 We believe the nucleolus is clearly the best choice that satisfies a majority of the above
 properties. Since nucleolus is defined as the unique utility vector which makes the
 unhappiest coalition as less unhappy as possible~\cite{straffin:book}, and given that
 the nucleolus is always in a non-empty core, it is the utility vector that retains the
 most number of bidders if the core is empty and is the most stable one retaining all the bidders if the core is non-empty.
 We compute the nucleolus by solving a series of linear programs~\cite{mwg,myerson:book} and
 obtain the utility vector $( x_1, x_2, \ldots, x_n)$ for the $n$ players and the search engine ($x_0$).


\subsection{Mapping the Utility Vector to a Correlated Bid Profile}\label{CorrBidProfileMapping}

\subsubsection{Obtaining a locally envy-free bid profile for each valid coalition}
To satisfy the stability criterion in the non-cooperative sense,
and ensure truthful participation of all the bidders in the
proposed bid optimizer, we aim to find out locally envy-free bids
for each of the $ n \choose k $ possible sets of winning bidders.
For finding these bids, consider a subset of $(k+1)$ bidders and
allocate slots to the bidders in this subset in the sorted order
of their willingness-to-pay values to satisfy the requirement for
the locally envy-free equilibrium. Now, the bids can be calculated
as follows. The $(k+1)^{th}$ bidder bids the reserve price
(assumed to be $0$ here without loss of generality). The bid of
the $k^{th}$ bidder (who pays $\overline{s}_{k+1}$) is now
calculated by solving for $b_k$ in $ \beta_k ( \overline{s}_k -
b_{k+1} ) = \beta_{k-1} ( \overline{s}_k - b_k) $ to satisfy the
envy-freeness. Once we obtain $b_k$, we proceed recursively by
replacing the $b_{k+1}$ by $b_k$ and $k$ by $(k-1)$ in the above
equation to get  $b_{k-1}$ and so on till we get the bids of all
the $k$ players. Note that the bid of the first player does not
have a role here as long as it is greater than the next highest
bid. Thus we obtain \textit{a} set of bids which are in locally
envy-free equilibrium.

\subsubsection{Obtaining a correlated bid profile}
The solution given by the nucleolus provides a utility for each
bidder. This cannot be used directly in the GSP auction of the
search engine. Towards this end, we map the nucleolus to a
correlated bid profile which defines the required rotation among
the bidders for occupying the slots. This correlated bid profile
is what is finally suggested by the bid optimizer, which retains
the maximum number of advertisers without hurting the search
engine.

The characterization of a correlated bid profile corresponds to
assigning the probabilities associated with each of the bid
profiles associated with the bidders. There exist several
algorithms in general, for finding the correlated bid profile. But
we would like to exploit the structure of the problem and obtain a
simpler solution without going into the complex details about
modifying the ellipsoid algorithm as done in most of the work in
this area. See ~\cite{papadi:correlated} for example. Any
correlated strategy we consider here has a subset of size $k$
bidders bidding their corresponding LEF (locally envy-free) bids
(obtained in the previous section) and all other bidders bidding
the reserve price. Considering only these $n \choose k$ strategy
profiles corresponding to each subset of size $k$ bidders winning
the slots would suffice since they exhaust all the possible
outcomes of the underlying GSP auction.

The probability distribution which yields the utilities suggested by the nucleolus to the players is any distribution which satisfies the constraints that it is a probability distribution, it is individually rational for each player and it must yield the payoffs suggested by the nucleolus to the bidders subject to their budget constraints. This can be obtained by solving a linear program as follows.

     $$ \min   \hspace{0.5 cm} \sum_{i \in N \cup \lbrace 0 \rbrace}  z_i + \sum_{i \in N} \overline{s}_i \left( x_{i} - \left( \sum_{\substack{C \subseteq N ,\> \mid C \mid = k \\ i \in C ,\> j = y_{iC} }} p_{C} \beta_j (\overline{s}_{i} - b_{j}) \right) \right) \hspace{5.6 cm}$$

 subject to

 \begin{eqnarray*}
 \forall i \in N \>\>\>\>\>\> z_i & \geq  & \left( \sum_{\substack{C \subseteq N ,\> \mid C \mid = k \\ i \in C ,\> j = y_{iC} }} p_{C} \beta_j (\overline{s}_{i} - b_{j}) \right) - x_{i}   \\
 \forall i \in N \>\>\>\>\>\> z_0  & \geq &  \left( \sum_{\substack{C \subseteq N ,\> \mid C \mid = k }} p_{C} \sum_{\substack{ i \in C ,\> j = y_{iC}}}\beta_j b_{j+1} \right) - x_0   \\
  \forall i \in  N \>\>\>\>\>\> z_0 & \geq  & x_0 - \left( \sum_{\substack{C \subseteq N ,\> \mid C \mid = k }} p_{C} \sum_{\substack{ i \in C ,\> j = y_{iC}}}\beta_j b_{j+1} \right)  \\
 \forall i \in N  \>\>\>\>\>\> z_i & \geq &  x_{i} - \left( \sum_{\substack{C \subseteq N ,\> \mid C \mid = k \\ i \in C ,\> j = y_{iC} }} p_{C} \beta_j (\overline{s}_{i} - b_{j}) \right) \\
 p_{C} \beta_j (\overline{s}_{i} - b_{j}) &  \geq  & 0 \>\>\>  \forall C \subseteq N \>\> \forall i \in C \>\> \forall j \in K \\
\sum_{C \subset N} p_{C} & = & 1  \\
\forall C \subset N \>\>\>\>\>\> p_{C} & \geq & 0
 \end{eqnarray*}
where $y_{iC} $ denotes the slot that player $i$ wins in a locally envy-free allocation if only the set $C$ of players were to win all the slots.

The linear program maps the utility vector suggested by the nucleolus into a correlated bid profile. The objective function minimizes the difference between the utility suggested by nucleolus and the expected utility in the correlated bid profile for each player. The minimization of difference leads to two constraints for each player. This is because for any two variables $x$ and $y$,
$$ \min \mid x - y \mid $$
is the same as
$$ \min z $$
$ \hspace{2 cm} $subject to
$$ z \geq x - y $$
$$ z \geq y - x $$

In the minimization, the higher valuation bidders are given a preference over the lower valuation bidders. This is done by weighting each player's difference from the nucleolus in the objective function by their valuation. This is a heuristic to ensure that the error in the inverse mapping of the utility vector to a correlated bid profile is biased towards the higher valuation bidders so that they voluntarily participate in the bid optimizer. Since the only problem to Individual rationality is when the higher valuation bidders shade their willingness-to-pay, this weighing gives the incentive for them to reveal their true valuations.

In the objective function, we minimize the \textit{difference} (this is done by the first $4$ constraints of the linear program) between the utility vector and the obtained expected utility in the above linear program since the restriction of the bid profiles to the set of locally envy-free equilibria may not have a feasible correlated bid profile. The minimization is done in such a way that the higher valuation bidders obtain relatively higher utility (due to the weights given to the difference in the objective function) than the lower valuation bidders in case the optimal value of the objective function is non-zero. This is a heuristic to ensure that the error in the inverse mapping of the utility vector to a correlated bid profile is biased towards the higher valuation bidders so they voluntarily participate in the bid optimizer.

\subsection{Properties of the Proposed Solution} \label{ANALYSIS}
%
The properties of the proposed solution are as follows:
\begin{itemize}
 \item The proposed solution has the bidders participating voluntarily in the bid optimizer for the following reasons. (i) The auctioneer is benefited since he has a guaranteed revenue of at least what is suggested by the nucleolus. (ii) The high valuation bidders are benefited since they are offered the same slots at a relatively lower price. Also, since the nucleolus tries to retain the grand coalition intact, it will be individually rational for the  high valuation bidders  to participate in the bid optimizer rather than to deviate and bid higher. (iii) The lower valuation bidders are benefited because they get more slots and hence more clicks and their campaign is more effective. Thus, the utility of every player increases and the individual rationality (IR) condition is satisfied.

 \item The bids suggested are in a locally envy-free equilibrium of the game and also are in the core since the nucleolus is always in core if the core is non-empty. This indicates that the proposed solution is strategically stable. In other words, no one can profitably deviate unilaterally from the solution proposed by the bid optimizer.

 \item Retention and efficient division of the worth of the grand coalition are guaranteed by the nucleolus since it is the allocation which tries to retain the grand coalition.

 \item Truthfulness is difficult to satisfy, given that the GSP mechanism is non-truthful. But note that the lower valuation bidders have no incentive to shade their bids. If they do so, they may lose their slots or run into negative utilities. Hence there is a problem only when the higher valuation bidders do not participate in the bid optimizer or they understate their willingness-to-pay. The higher valuation bidders cannot understate their valuations by a large amount since they have a threat of losing their slots to lower valuation bidders who are retained in the system. Also, the higher valuation bidders are given more benefits to participate in the bid optimizer and it is individually rational for them to participate in the bid optimizer.
\end{itemize}
 Hence this solution satisfies all the properties which were mentioned in Section \ref{IdealProperties}.


\section{Experimental Results}\label{EXPERIMENTS}
This section presents simulation based experimental results to explore the effectiveness of the approach presented in the paper. First we start with a model for the drop outs of bidders.

\subsection{Bidder drop out model}
The bidders drop out if they do not get enough slots (or alternatively clicks) consistently over a period of time. The conditional probability that a bidder drops out given that he did not get a slot in a round may vary from bidder to bidder. Also, the positions of slots occupied by the bidders in the previous few rounds of auction could play an important role in the dropping out of a bidder.

To model this behavior of the bidder dropping out based on the outcomes of the previous auctions and giving more importance to recent outcomes, we propose a discounted weighting of the outcomes of the previous auctions to compute the probability that a bidder will continue in the next round of the auction. This model also captures the myopic human behavior that the bidder's choice is dependent on only the recent outcomes. That is, the history the bidder looks into, before taking a decision to continue or not for the next round is limited. The amount of history however depends on the bidder in the form of his discounting factor. Let $x_{-i} \in \lbrace 0, 1 \rbrace$ denote the outcome of the  $i^{th}$ previous round. A ``$1$`` denotes that the bidder received a click in the $i^{th}$ previous auction with a zero indicating otherwise. We propose that the probability that the bidder will participate in the next round is given by $\frac{\sum_{i=1}^{\infty} \gamma^i x_{-i}}{\sum_{i=1}^{\infty} \gamma^i} = (1 - \gamma) \sum_{i=1}^{\infty} \gamma^ix_{-i}$ where $\gamma$ is the discount factor of the bidder. To see how the myopic nature of the bidders is captured, suppose that the bidder's discount factor $\gamma = 0.95$. The discount factor for the $101^{st}$ round will be $0.006$ which is negligible. Hence the bidder's decision is dependent on at most $100$ previous auctions. Thus, the discount factor decides the nature of the bidder.

\subsection{Experimental Setup}
Given a fixed set of CTRs, the valuations of the bidders are chosen close enough to each other, to analyse our model in competitive environment. The retention problem, is fundamental in competitive environment, as search engine needs to retain the bidders and allow them compete in further auctions. The results indicate that the proposed bid optimizer not only retains a higher number of bidders than the normal GSP but also yields better cumulative revenue to the search engine in the long run.

\input{epsf}
\begin{figure}[!hbtp]
\hspace{-2cm}

\epsfxsize=9 cm
\epsfysize=7 cm
 \epsfbox{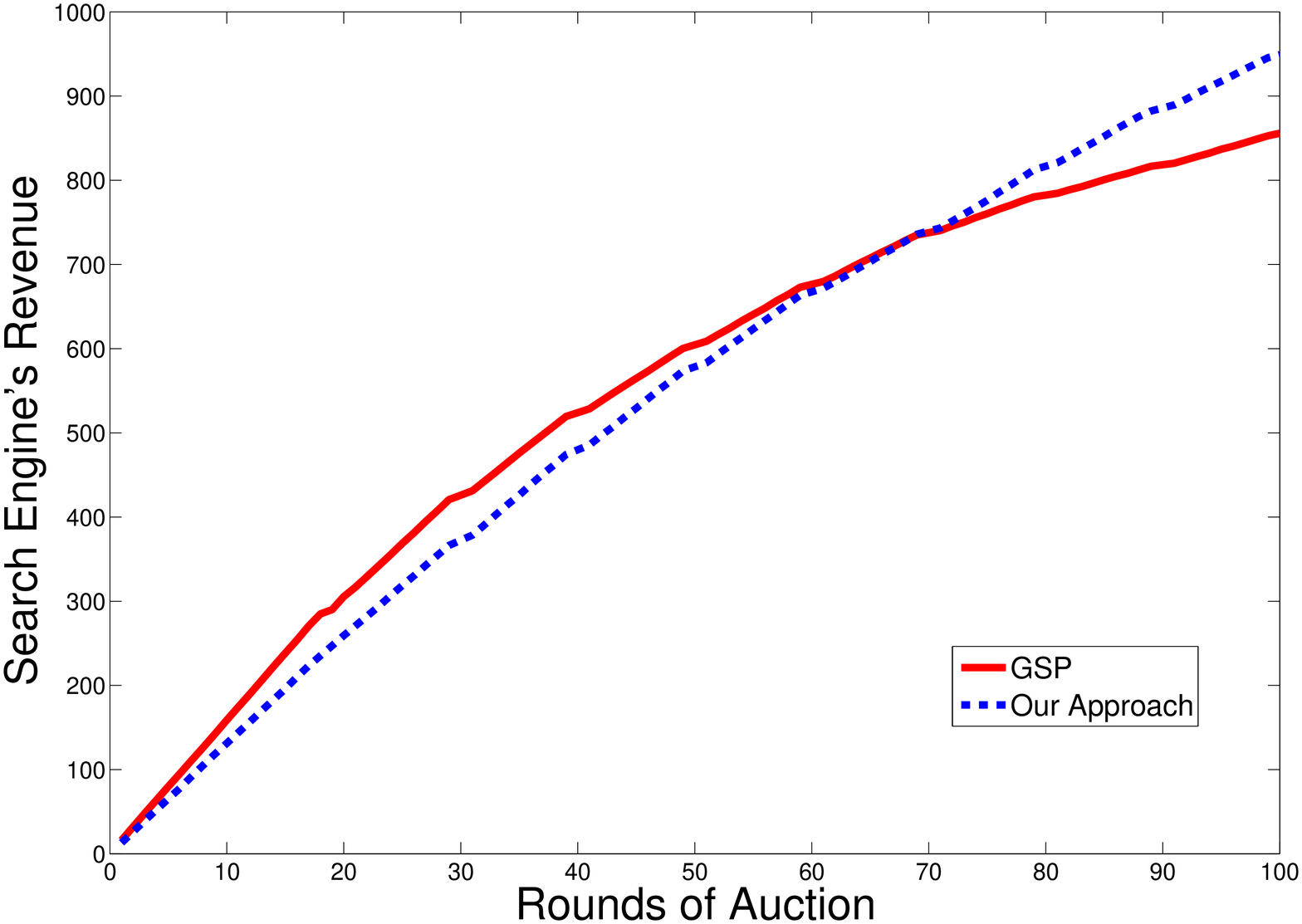}
 \caption{Cumulative revenue of the search engine}
 \label{fig:exp1}
\end{figure}

\subsection{Cumulative Revenue of the Search Engine}

First we consider the cumulative revenue of the search engine for
comparing the non-cooperative bidding and using cooperative bid
optimizer. We consider $10$ advertisers with $5$ slots to be
allocated.We run successive auctions and find the cumulative revenue of the search
engine after each auction using the two approaches. We run this experiment until
the change in the average cumulative revenue after each auction
becomes acceptably small. Figure \ref{fig:exp1} shows a comparison  of the cumulative revenue of the search engine under the proposed approach with that of a standard GSP auction.

It can be seen in Figure \ref{fig:exp1} that though initially the
non-cooperative approach (GSP) yields more revenue, after a few runs,
the cooperative bid optimizer, with all the solution vectors,
starts outperforming. Initially the GSP outcome is
better, as the advertisers are bidding their maximum willingness
to pay, and hence the search engine gets high levels of revenue. However, as
the utilities of the advertisers are less in the case of the
non-cooperative approach, they start dropping out of the auction
and hence in a long run, the cumulative revenue starts declining
compared to the cooperative bid optimizer. This is the adverse
effect of the dropping out of the advertisers leading to the
retention problem.
\input{epsf}
\begin{figure}[!hbtp]
\hspace{-1cm}
\epsfxsize=10 cm
\epsfysize=8 cm
 \epsfbox{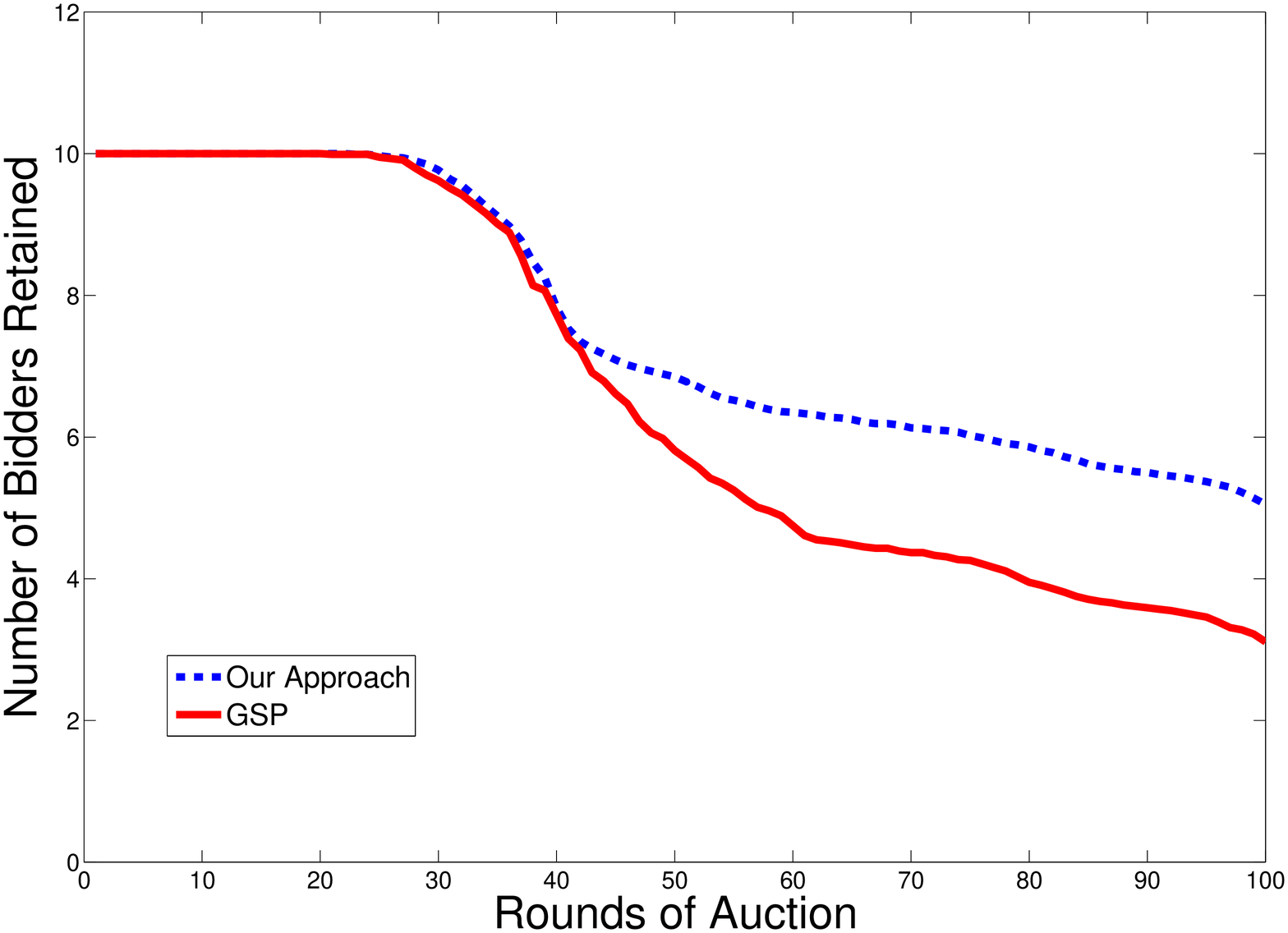}
 \caption{Number of bidders retained in the system}
 \label{fig:exp3}
\end{figure}
\subsection{Number of Bidders Retained in the System}

After each auction, we compute the average number of advertisers
retained, to analyze the  retention dynamics in the system.
Figure \ref{fig:exp3} gives a comparison between
using the cooperative bid optimizer and the GSP approach.

In Figure \ref{fig:exp3}, it can be observed that there are
considerably more number of advertisers retained in the system
when using the solution vector suggested by the cooperative bid
optimizer in comparison to the GSP approach. This
explains the reason for the dip in the non-cooperative
cumulative revenue of the search engine.

Through experimentation, we are able to  demonstrate revenue increase for the
search engine in the long run and also reduction of the retention
problem considerably. It should be noted that even though the
raise is not substantial, its impact on retaining and attracting the
advertisers and thereby other indirect advantages are immense.

\section{Summary and Future Work} \label{SUMMARY}
We have proposed a bid optimizer for sponsored keyword search auctions which leads to better retention of advertisers by yielding attractive utilities to the advertisers without decreasing the long-run revenue to the search engine. The bid optimizer is a value added tool the search engine provides to the advertisers which transforms the reported values of the advertisers for a keyword into a correlated bid profile. The correlated bid profile that we determine is a locally  envy-free equilibrium and also a correlated equilibrium of the underlying game. Through detailed simulation experiments, we have shown that the proposed bid optimizer retains more customers than a plain GSP mechanism and also yields better long-run utilities to the search engine and the advertisers.

The experiments were carried out with a model that captures  the phenomenon of customer
drop outs and showed that our approach produces a better long run
utility to the search engine and all the advertisers. The proposed bid optimizer is beneficial for both the bidders and the search engine in the long run.

We considered GSP auction, which is popularly run in most of
the search engines, in our analysis. However, it would be
interesting to look at the effects of other auction mechanisms
like VCG auctions on the overall process.

The other important components like budget optimization and
ad scheduling are involved in  sponsored search
auctions. We would further like to combine these components with
our cooperative bid optimizer.

\section*{Appendix}
Nash~\cite{Nash} proposed that there exists a unique solution
function $f(F,v)$ for every two person bargaining problem, that
satisfies the following 5 axioms - \em{Pareto strong efficient,
Individual Rationality, Symmetry, Scale Covariance, and
Independence of Irrelevant Alternatives.} The solution function is
$$  f(F, v) \in {\rm argmax}_{ (x_1,x_2) \in F}
((x_1-v_1)(x_2-v_2))$$

where, $x_1 \geq v_1$ and $x_2 \geq v_2$
and the point $v = (v_1, v_2)$ is known as the \em{point of
disagreement}. There are several possibilities for choosing the
disagreement point $v$. The three popular choices are those based
on (1) a minimax criterion, (2) focal equilibrium, and (3)
rational threats. As part of this paper, we use the rational
threats to identify disagreement point $v$~\cite{myerson:book}.
For more details please refer to the books~\cite{myerson:book}
~\cite{straffin:book}.

\end{document}